\documentclass[final,5p,times,twocolumn]{elsarticle}
\usepackage{lineno,hyperref}
\usepackage{amsfonts}
\usepackage{amsmath}
\usepackage{amssymb}
\usepackage{latexsym}
\usepackage{color}
\usepackage{graphicx,amssymb,bm,makeidx,amsmath}
\modulolinenumbers[5]

\journal{Journal of \LaTeX\ Templates}

\renewcommand{\k}{\mathbf{k}}

\begin{document}

\begin{frontmatter}

\title{Giant density-of-states van Hove singularities in the face-centered cubic lattice}

\author{P.A. Igoshev}
%\address{IMP, Ekaterinburg, Russia}


\author{V.Yu. Irkhin}
\address{M. N. Mikheev Institute of Metal Physics, 620108 Ekaterinburg, Russia}

\begin{abstract}
All van Hove singularities in the density of states (DOS) of face-centered cubic lattice in the nearest and next-nearest neighbour approximation, focusing on higher-order ones,  are found and classified. At special values of the ratio $\tau$ of nearest ($t$) and next-nearest neighbour hopping integral $t'$ giant DOS singularities, caused by van Hove lines or surfaces,  are formed. An exact formula for DOS which provides  efficient numerical implementation is proposed. The standard tetrahedron method is demonstrated to be inapplicable due to its poor convergence in the vicinity 
%of energy positions 
of kinks caused by van Hove singularities. A comparison with the case of large space dimensionality (infinite coordination number) including next-nearest neighbours 
%with a scaling providing a balance of nearest- and next-nearest contributions into local state decaying  
%DMFT approach 
is performed. 
%The contribution to DOS of  isolated  heavy-mass van Hove points is considered too. 
%??The impact of the giant singularities onto physical properties (heat capacity, susceptibility, etc.) is discussed. 
\end{abstract}


\end{frontmatter}

%\linenumbers

%\section{The Elsevier article class}

\section{Introduction}
Zero-velocity $\mathbf{k}$ points in the Brillouin zone are inevitably present in the  one-particle excitation spectrum of a periodic system, as follows from the  Morse theorem in differential topology (see~\cite{1953:vanHove}). 
A peculiar role of such points is manifested by the fact that each such $\mathbf{k}$-point generally produces a divergence in the \textit{derivative} of the corresponding density of states (DOS) at the  energy for this $\mathbf{k}$. We refer to such a DOS feature as the van Hove singularity  (vHS)~\cite{1953:vanHove}. This singularity can be enhanced by some factors (e.~g.~merging of several singularities, effective-mass divergence, etc.), which results in the DOS divergence at the corresponding energy, leading  to significant effects in physical properties. 

The  vHS's play an important role in the solid state physics leading to important consequences for electronic, magnetic and structural properties. Formation of strong van Hove singularities results in the localization of the electron states and in an increase in the role of correlation effects~\cite{1993:Vonsovskii}. Therefore these singularities, especially electronic DOS peaks near the Fermi level, are crucial for ferromagnetic ordering~\cite{1938:Stoner,1968:Edwards} and favor other Fermi-liquid instabilities, including superconductivity (see, e.g., \cite{Park}). 
A geometrical origin of DOS peaks in a general context, as well as its implementation for concrete compound electronic spectrum obtained within the density functional theory (DFT) were investigated in detail in~Ref.~\cite{1990:Peschanskikh}. In particular, in fcc phase of Sr, weakly dispersive one-dimension manifolds (their prototype is a van Hove line) are positioned at the face of the Brillouin zone near the Fermi level. They correspond to X--U, U--L, L--K, K--U, and K--W directions. %?? other examples

Recently, so-called higher-order van Hove singularities  connected with divergence of some effective masses have attracted a lot of attention, in particular as a possible cause of unconventional insulating state and superconductivity in twisted bilayer graphene at a magic angle~\cite{2019:Yuan,2018:Isobe,2019:Kozii}. A giant vHS was found within DFT calculations~\cite{2021:Hung} in Mg$_3$Bi$_2$ compound with high thermoelectric power. Non-trivial  consequences of electron spectrum features were found in nodal-line semimetals: three-dimensional ZrSiSe~\cite{2020:Shao} and bilayer ZrSiS~\cite{2018:Rudenko} possessing  a whole vHS line. 
We state that peculiarities of spectrum may have various and valuable impact onto physical properties and its isoenergetic surface topology should be investigated carefully within numerical and analytical aspects. 

Van Hove singularities are also extensively discussed now in the context of the so-called Hund's metals where they can result (together with asymmetry of DOS with respect to the Fermi level) in a considerable enhancement of quasiparticle damping (up to destroying Landau quasiparticle behaviour) as a consequence of the Hund coupling~\cite{2018:Belozerov}.  Compounds with particular narrow bands can serve as examples of such systems: $\alpha$- and $\gamma$-Fe, and Ni were considered as candidates to the Hund's metal group~\cite{2018:Belozerov}. 
The Hund-metal features are similar to the orbital-selective Mott transitions, introduced  to explain unusual properties of Ca$_{2-x}$Sr$_x$RuO$_4$~\cite{2002:Anisimov}, which are to a large extent determined by vHS. 
Strong and non-monotonous temperature behaviour of magnetic susceptibility in LaFeAsO is caused by a~closeness of~the DOS peak, possibly produced  by van Hove singularity near the Fermi level~\cite{2011:Skornyakov}. 
Therefore it is instructive and useful for concrete calculations to have explicit DOS expressions. 

Early investigation~\cite{1969:Jelitto} presented a numerical interpolation of DOS for cubic lattices, but only the case of nearest-neighbor hopping approximation (integral $t$) was considered. At the same time, for modeling real compounds the next-nearest-neighbor hopping $t'$ plays  often  an important role. 
For the square lattice, this point was widely discussed in the context of superconducting cuprates  within  the non-degenerate Hubbard model. Here, simple N\'eel antiferromagnetic order, being present in the nearest-neighbour approximation  at half-filling for  arbitrarily small Hubbard's $U$, can be replaced by incommensurate magnetic ordering~\cite{2010:Igoshev,2015:Igoshev}, $d$-wave susceptibility~\cite{2003:Kampf}, ferromagnetic ordering~\cite{2010:Igoshev,2015:Igoshev,2003:Kampf,Hlubina}, provided that considerable next-nearest-neighbour hopping is introduced. 

The van Hove singularity owing to L~point in the fcc lattice Brillouin zone 
produces weak ferromagnetic ordering in ZrZn$_2$ and plays important role in unusual  physical properties: metamagnetic transition in a weak  magnetic field, Lifshitz transitions, non-Fermi-liquid behaviour~\cite{2003:ZrZn2:Yates,2004:ZrZn2:Major,2007:ZrZn2:Amaji,2012:ZrZn2:Kabeya}.  Magnetic ordering in Ni is mainly caused by the vicinity of L~point to~the~Fermi surface  
%of Brillouin zone 
(so-called ``van Hove magnet'')~\cite{2017:Ni:vanHoveMagnet}, which dramatically distinguishes this from  systems with well-formed local moments like $\alpha$-Fe and rare-earth systems. 

From this point of view, the case of three-dimensional bipartite lattices (simple cubic and body-centered cubic) was investigated in~Refs.~\cite{2019:Igoshev-PMM,2019:Igoshev-JETP}, closed analytical expressions for DOS as a function of energy and $\tau = t'/t$ in terms of elliptic integrals being obtained.

For a non-bipartite fcc lattice, an analytical investigation of the Van Hove energy levels and numerical analysis of the DOS within the nearest and the next-nearest neighbour hopping approximation were performed to a great extent only in a limited interval $0< \tau < +1$~\cite{1972:Swendsen}, which did not allow to find van Hove singularity lines. 
At the special value $\tau  = -1/2$, a giant Van Hove line occurs at the band edge producing a DOS singularity of the inverse square-root type (for the case of infinite-dimension lattice it was discussed in~Ref.~\cite{Ulmke}). Here we provide an analysis of stability of this singularity and of its influence on %thermodynamic properties, as well as 
the  DOS picture  in the whole $\tau$ interval.

\section{Electron spectrum and giant van Hove singularities at $\tau = -1/2, 0, 1$}
We write down electron spectrum as
%in the nearest and next-nearest neighbor approximation as
\begin{multline}\label{eq:spectrum}
t_{\rm fcc}\left(\mathbf{k};\tau\right) = -4 t\left(  \cos{k_x}\cos{k_y} + \cos{k_x}\cos{k_z}+\cos{k_y}\cos{k_z} \right)
\\+2t\tau\left(\cos 2  {k_x} +\cos 2{k_y} +\cos2  {k_z} \right),
\end{multline}
where we assume nearest-neighbour hopping integral $t > 0$ and parametrize the next-nearest-neighbour hopping integral $t' = \tau t$~(lattice parameter is chosen as $a = 2$). Below we take $t = 1$ for brevity. The definition of nearest-neighbor hopping integral is the same as in Ref.~\cite{1969:Jelitto}. Density of states is defined by the following integral over Brillouin zone
\begin{equation}\label{eq:DOS_fcc_def}
    \rho_{\rm fcc}(\epsilon, \tau = 0) = \int \frac{d^3\mathbf{k}}{V_{\rm BZ}}\delta(\epsilon - t_{\rm fcc}\left(\mathbf{k};\tau\right)).
\end{equation}

\begin{figure}[t!]
%\noindent
\centering
\includegraphics[width=0.45\textwidth]{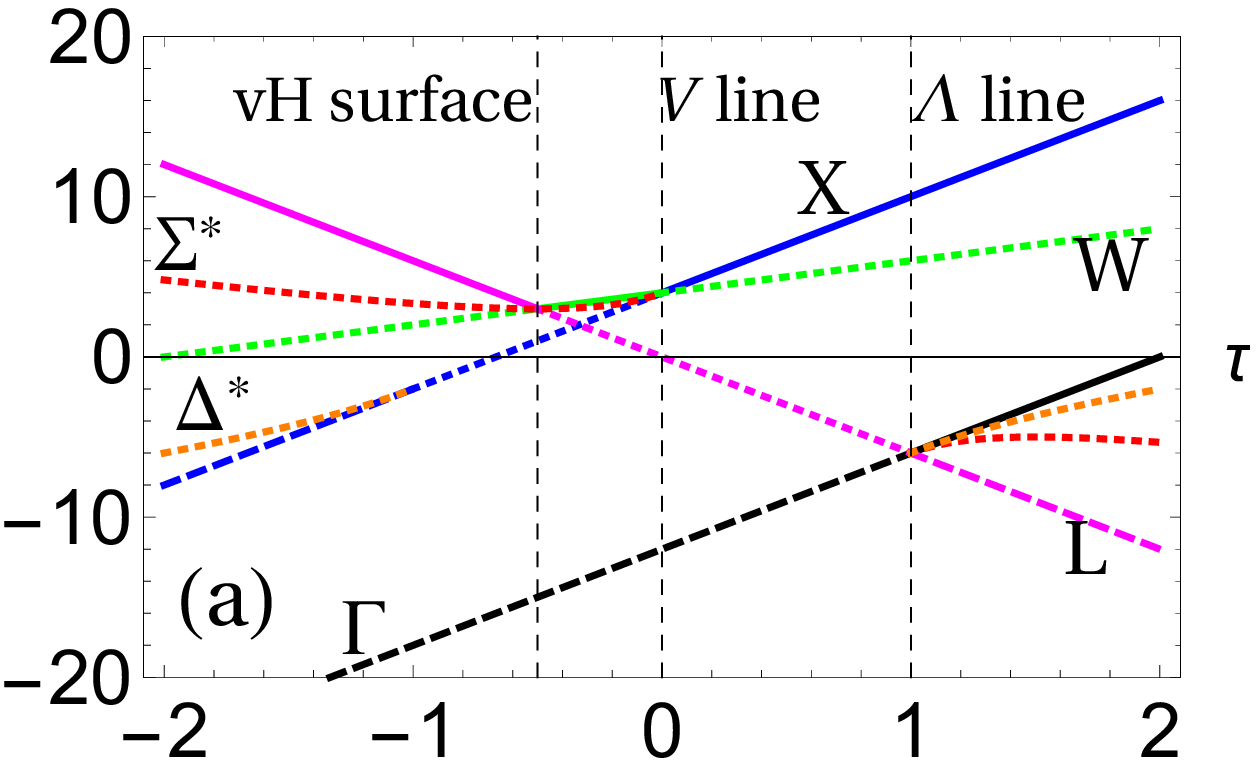}
\includegraphics[angle=-90,width=0.35\textwidth]{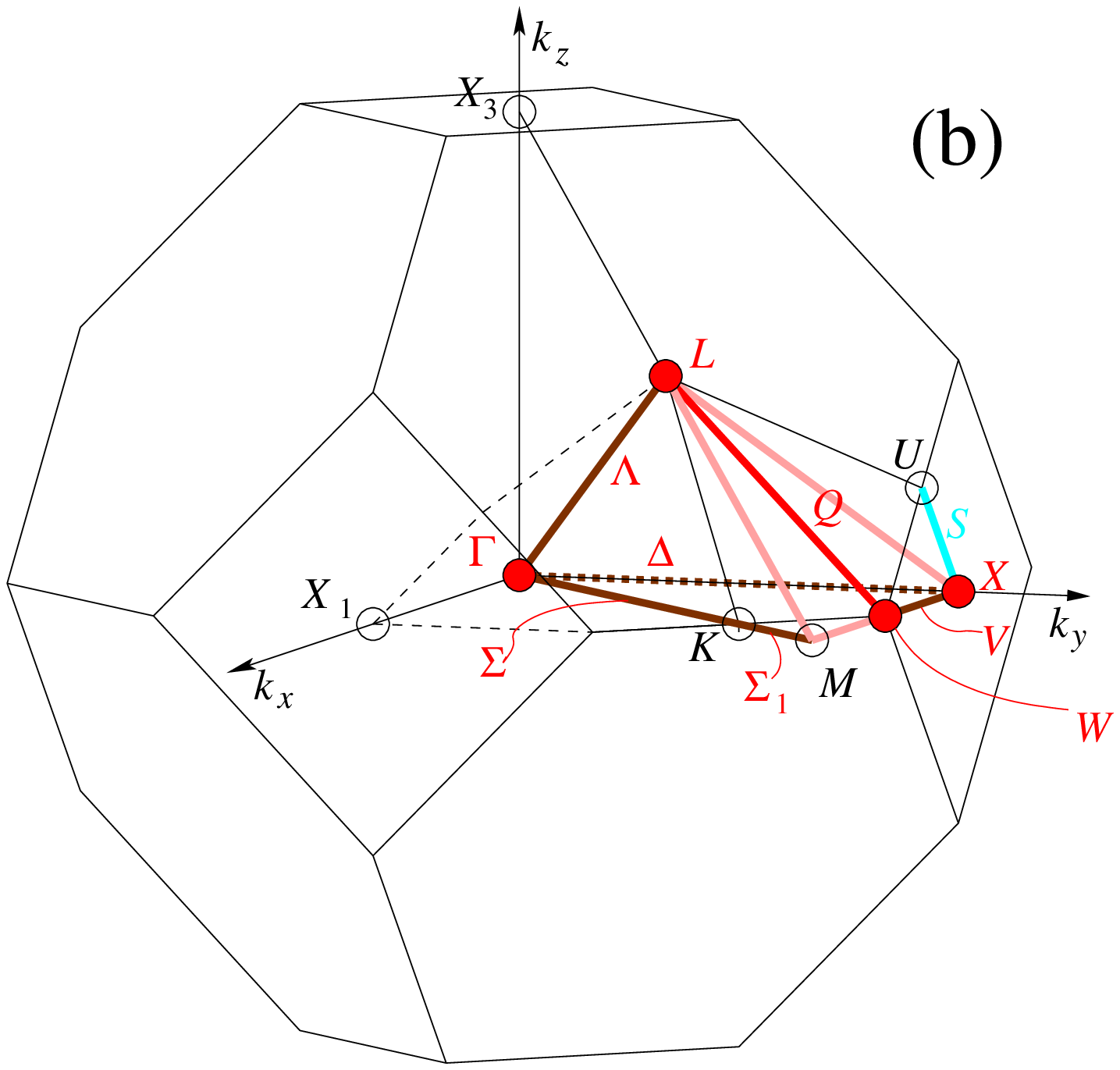}
\caption{(a) $\tau$ dependence of the energy $w$ of van Hove $\mathbf{k}$ points for fcc lattice, see definitions in Table~\ref{table:fcc}. 
Solid line denotes a local maximum point, dashed line  a local minimum point, dotted line line  a saddle-point. (b) Brillouin zone of fcc lattice with notations for high-symmetry points and directions. 
\label{fig:w_fcc}
}
\end{figure}

In~Table~\ref{table:fcc}  we present an analysis of van Hove $\k$ points  in the vicinity of van Hove $\mathbf{k}$ point) depending on $\tau$. 
We write down  their energy and  \textit{signature}, i.e. the difference of positive and negative indices of inertia of the second order $t_{\rm fcc}\left(\mathbf{k};\tau\right)$ Taylor-expansion terms.  
For further convenience, an excerpt from this in a demonstrative manner is given in~Fig.~\ref{fig:w_fcc}. 

An analogous Table I is presented in Ref.~\cite{1972:Swendsen}, being restricted to $\tau < 1$ ($\gamma =-\tau$ in that consideration).  To compare, $0$ point of~\cite{1972:Swendsen} corresponds to $\Gamma$ being a global minimum at $\tau < 1$ and a local maximum at $\tau > 1$; A point of~\cite{1972:Swendsen} corresponds to L and actually changes its type at $\tau = +1$ to global minimum at $\tau > 1$; B point of~\cite{1972:Swendsen} corresponds to W and does not change its vHS type at $\tau = -1$, as is claimed in Ref.~\cite{1972:Swendsen}; C point of~\cite{1972:Swendsen} corresponds to X; D point of Ref.~\cite{1972:Swendsen} corresponds to $\varDelta^\ast$ and is realized at $|\tau| > 1$ with /+1/ at $\tau < -1$ and /-1/ at $\tau > +1$, which is not taken into account in \cite{1972:Swendsen} (also the expression for the energy contains a typo); E point of~\cite{1972:Swendsen} corresponds to $\varSigma^\ast$. Thus the Table I of Ref.~\cite{1972:Swendsen} contains some errors and typos.

\begin{table*}[t!]
\caption{\label{table:fcc}Non-equivalent $\k$ points of van Hove singularities for FCC lattice, see.~Fig.~\ref{fig:w_fcc}. Arrow denotes a change of signature of mass tensor of van Hove $\mathbf{k}$ point as  $\tau$ increases above the reference $\tau_\ast$ value. $k_{\varSigma^\ast} =  \arccos(2\tau - 1)^{-1}, k_{\varDelta^\ast} = \arccos\tau^{-1}$. $a^{(1)}_{\rm \varDelta^\ast} = 4(\tau^{-1}-1)(1 + 2\tau)$, $a^{(2)}_{\rm \varDelta^\ast} = 8\tau^{-1}(\tau^2 - 1)$, $a^{(1)}_{\rm \varSigma^\ast} = 8(\tau - 1)(1 + 2\tau)/(1 - 2\tau)$,	$a^{(2)}_{\rm \varSigma^\ast} = 16\tau(\tau - 1)(1 + 2\tau)(1 - 2\tau)^{-2}$, $a^{(3)}_{\rm \varSigma^\ast} = 16\tau(\tau - 1)(2\tau - 1)^{-1}$. We distinguish four types of van Hove points: minimum (``min''), maximum(``max''), $/\pm1/$ corresponds to signature of the Hessian matrix of $t(\k;\tau')$ at the saddle van Hove point $\k = \k_{\rm vH}$. $\varSigma^\ast (\varDelta^\ast)$ denotes van Hove point located at high-symmetry Brillouin zone direction $\varSigma (\varDelta)$.
}
\center
\begin{tabular}{|c|c|c|c|}
\hline
$\mathbf{k}$ & $w = t(\mathbf{k})$ & inverse masses & signature \\
\hline
$\Gamma(0,0,0)$ & $-12 + 6\tau$ & $8(1 - \tau),8(1 - \tau),8(1 - \tau)$&min$\stackrel{\tau = 1}{\rightarrow}$max\\
\hline
X$(0,0,\pi)$ & $+4 + 6\tau$ & $-8\tau,-8\tau, -8(1+\tau)$&min$\stackrel{\tau = -1}{\rightarrow}/+1/\stackrel{\tau = 0}{\rightarrow}{\rm max}$\\
\hline
W$(0,\pi/2,\pi)$ & $+4 + 2\tau$ &$-4(1 + 2\tau),8\tau,8\tau$&$/+1/\stackrel{\tau = -1/2}{\rightarrow}{\rm max}\stackrel{\tau = 0}{\rightarrow}/-1/$\\
\hline
L$(\pi/2,\pi/2,\pi/2)$ & $-6\tau$ & $ 4(1 + 2\tau), 4(1 + 2\tau),8(\tau - 1)$&${\rm max}\stackrel{\tau = -1/2}{\rightarrow}/+1/\stackrel{\tau = +1}{\rightarrow}{\rm min}$\\
\hline
${\varDelta^\ast}\left(0,0, k_{\varDelta^\ast}\right)$ & $-2\tau^{-1}(2 + 2\tau - \tau^2)$ & $a^{(1)}_{\rm \varDelta^\ast},a^{(1)}_{\rm \varDelta^\ast},a^{(2)}_{\rm \varDelta^\ast}$ & $/-1/, \tau > +1$; $/+1/, \tau < -1$ \\
\hline
${\varSigma^\ast}\left(0,k_{\varSigma^\ast},k_{\varSigma^\ast}\right)$ & $4(1 - 2\tau)^{-1}-2\tau$& $a^{(1)}_{\rm \varSigma^\ast},a^{(2)}_{\rm \varSigma^\ast},a^{(3)}_{\rm \varSigma^\ast}$ & $/+1/, \tau > +1$; $/-1/, \tau < 0$\\
\hline
\end{tabular}
\end{table*}

The $\tau$ dependence of van Hove energies presented in Fig.~\ref{fig:w_fcc} implies that at special values $\tau = -1, -1/2, 0, 1/2, 1$ isoenergetic surfaces demonstrate topological transitions (change of van Hove singularity types).

The derivation of DOS formula for fcc lattice is presented in the next Section, see~Eq.~(\ref{eq:rho_zeta_final_mod}--\ref{eq:R_def}). Below in this Section we present only  calculation results and qualitative analysis. %\comPI{See!}
We have three special cases providing van Hove singularities: $\tau = -1/2$, $\tau = 0$ and  $\tau = 1$. 
For $\tau = -1/2$ the whole van Hove surface 
\begin{equation}
\cos k_x + \cos k_y + \cos k_z = 0 
\end{equation}
is formed, 
%The energy of this surface 
which corresponds to  merging of energy levels for W, L and $\varSigma^\ast$~points,
%for which $\cos k_x + \cos k_y + \cos k_z = 0$ corresponds to van Hove singularity surface
and results in giant quasi-one-dimensional behavior. However in this case DOS can be calculated easily, since the spectrum can be represented as $t_{\rm fcc}\left(\mathbf{k};-1/2\right) = 3 - t^2_{\rm sc}(\mathbf{k})/2$, where $t_{\rm sc}(\mathbf{k}) = 2(\cos k_x + \cos k_y + \cos k_z)$ is tight-binding spectrum for the simple cubic lattice in~the~nearest-neighbour approximation. 
Direct use of this formula implies
\begin{equation}
	\rho_{\rm fcc}(\epsilon; \tau = -1/2) = \frac{\sqrt{2}\rho_{\rm sc}(\sqrt{2(3 - \epsilon)})}{\sqrt{3 - \epsilon}}.
\end{equation}
with $\rho_{\rm sc}(\epsilon)$ being the corresponding simple cubic lattice DOS. This implies that van Hove singularity exists at the top of the band $\epsilon = +3$, and in the vicinity of it we have the following ``one-dimensional'' asymptotics 
$\rho_{\rm fcc}(\epsilon) = \sqrt2\rho_{\rm sc}(0)/\sqrt{3 - \epsilon} + \mathcal{O}(\sqrt{3 - \epsilon})$. 
Such a singularity is analogous to that for the square lattice in the next-nearest-neighbour hopping approximation at $t'=0.5t$ when the flat band in the vicinity of van Hove lines $k_x = 0$ and $k_y = 0$ is formed~\cite{2017:Igoshev}. 
%\comPI{Here we can be more exact taking into account subleading terms of expansion $\rho_{\rm sc}(\epsilon) = \rho_{\rm sc}(0) + A_{\rm sc}\epsilon^2 + \ldots$!} 

\begin{figure}[t!]
\noindent
\includegraphics[angle = -90, width=0.45\textwidth]{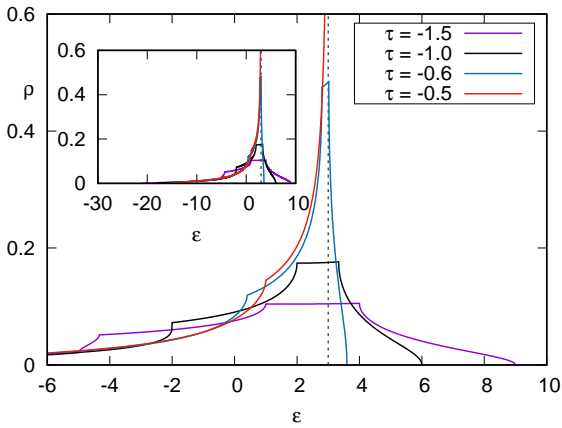}
\caption{DOS of fcc lattice calculated using numerical integration (\ref{eq:rho_zeta_final_mod}) at $\tau = -1.5, -1, -0.6$, and $-1/2$ within the region in the vicinity of giant van Hove singularity level $\epsilon = +3$ at $\tau = -1/2$ (shown by dashed vertical line). DOS plot in full energy range is shown in the inset.  Values of $\tau$ are shown in the legend. 
}
\label{fig:dos_fcc_tau1}
\end{figure}

\begin{figure}[t!]
\noindent
\includegraphics[angle = -90, width=0.45\textwidth]{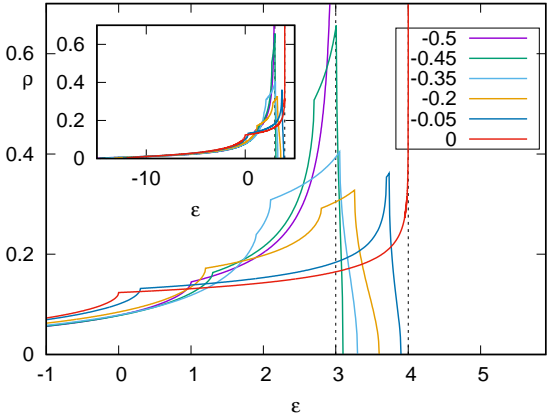}
\caption{The same as Fig.~\ref{fig:dos_fcc_tau1} for $\tau = -1/2, -0.45, -0.35, -0.2, -0.05$, and $0$ within the region in the vicinity of giant van Hove singularities level $\epsilon = +4$ (at $\tau = 0$) and $\epsilon = +3$ (at $\tau = -1/2$) (shown by dashed vertical line). DOS plot in full energy range is shown in the inset.
\label{fig:dos_fcc_tau2}
}
\end{figure}

From Figs. 2--5, one can  trace the influence of small deviations from critical $\tau$ values on DOS  plots. 
We deal with giant van Hove singularities of three types: (i) the logarithmic one (for $\tau = 0$)  at the top of the band; this was found by R.~Jelitto ~\cite{1969:Jelitto}; (ii) the inverse fourth-order-root type which was found at $\tau = 1$ at the bottom of the band; (iii)  quasi-one-dimensional one of inverse-square-root type ($\tau = -1/2$)  at the top of the band, which originates from van Hove surface containing high-order van Hove points. From Fig.~\ref{fig:dos_fcc_tau2} one can see that logarithmic singularity is much weaker than inverse-square-root type singularity. This relation holds under deviation of $\tau$ from critical values corresponding to topological transitions. At the same time, the difference of (ii) and (iii) singularities is practically not seen on the scale shown.

The transition of  $\tau$ through each special point  $\tau = -1/2, 0, 1$ causes a divergence of one or more of electron masses, see Table~1. This means that at deviation of $\tau$ from the ``critical'' value the van Hove line or surface decays into a few points with heavy masses. 

At $\tau \lesssim -1/2$, see~Fig.~\ref{fig:dos_fcc_tau1}, two van Hove points~W$/+1/$ and $\varSigma^\ast/-1/$  are formed with close energies,  $w^{\rm fcc}_{\varSigma^\ast}(\tau) - w^{\rm fcc}_{\rm W}(\tau)\sim-(1 + 2\tau)$; there is a narrow plateau between these energy levels. 
At the same time the occurrence of a kink at van Hove energy $w_{\varDelta^\ast}$ at $\tau < -1$ does not change substantially the DOS form.  

\begin{figure}[t!]
\noindent
\includegraphics[angle = -90, width=0.45\textwidth]{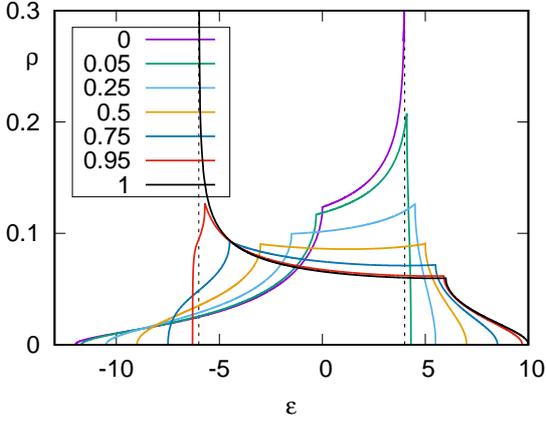}
\caption{DOS plot for $\tau = 0, 0.05, +0.25, +0.5, +0.75, 0.95$, and~$1$. 
\label{fig:dos_fcc_tau3}
}
\end{figure}

For $\tau$  larger than $-1/2$ (up to $\tau = 0$), the energy interval between these two points near the band top corresponds %and becomes corresponding
%A pair of these levels corresponds 
to strong falling of DOS from 
the  energy of the saddle point ($\varSigma^\ast$) to the energy of the maximum (W) with heavy mass at $\tau$ close to 0 and~$-1/2$, see Fig.~\ref{fig:dos_fcc_tau2}. According to Table~\ref{table:fcc}, internal van Hove points $\varDelta^\ast$ and $\varSigma^\ast$ are absent in the case~$|\tau| \le 1$. 
The least interesting case  $0 < \tau < 1$ was  investigated in detail in Ref.~\cite{1972:Swendsen}, see  Fig.~\ref{fig:dos_fcc_tau3}. 
Inside this interval, there are no internal van Hove points, as well as  van Hove lines and surfaces but at the ends of this interval, $\tau = 0, 1$ giant van Hove singularities occur. 

\begin{figure}[t!]
\noindent
\includegraphics[angle = -90, width=0.45\textwidth]{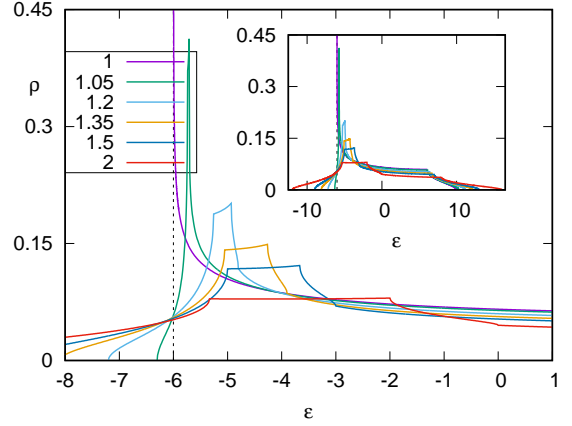}
\caption{The same as Fig.~\ref{fig:dos_fcc_tau1} for $\tau = 1, 1.05, 1.2, 1.35, 1.5$, and $2$. 
\label{fig:dos_fcc_tau4}
}
\end{figure}

At $\tau \gtrsim +1$, the van Hove line $\varLambda$ present at $\tau = +1$ and producing giant van Hove singularity decays into three van Hove points: $\varDelta^\ast, \varSigma^\ast,\Gamma$. Besides that, the energies of two first points form a typical narrow plateau
% a narrow plateau is formed??
with the width $ w^{\rm fcc}_{\varDelta^\ast}(\tau) - w^{\rm fcc}_{\varSigma^\ast}(\tau)\sim (\tau - 1)^2$ between levels of opposite-signature saddle points, see Fig~\ref{fig:dos_fcc_tau4}.

For  $\tau = 0$, the van Hove line  $V = \{(0,\pi,k_z), -\pi/2<k_z<\pi/2\}$ joining W and X points corresponds to merging of energy levels of these points. Therefore we get for inverse transverse masses $m^{-1}_{V}(k_z) = -4(1 \pm \cos k_z)$.
In this case, we represent the spectrum in the vicinity of the $V$ line as
%\begin{multline}\label{eq:t_fcc_tau=0_expansion}
%\end{multline}
\begin{multline}\label{eq:delta_rho.tau=1}
  E _V(k_z,q_x,q_y) = t_{\rm fcc}\left(\mathbf{k}_V(q_x, \pi + q_y, k_z);\tau = 0\right) = 4 - 2(1 - \cos k_z)q_x^2 \\- 2(1 + \cos k_z)q_y^2  
   + (1 - \cos k_z)q_x^4/6  + (1 + \cos k_z)q_y^4/6  + q_x^2q_y^2.  \nonumber
\end{multline}
Expanding the spectrum in $k_z = q_z$ we get  up to leading  %fourth-order 
terms 
$$t_{\rm fcc}\left(\mathbf{k}_V(q_x, \pi + q_y, q_z);\tau = 0\right) = 4 - 4q^2_y - q^2_xq^2_z$$ 
which implies that X is a higher-order van Hove point, since only one inverse mass is non-zero. 


We have six non-equivalent $V$ lines, each pair of them being positioned at the square faces of the Brillouin zone. 
The contribution of the vicinity of these lines, see~Eq.~(\ref{eq:DOS_fcc_def}),  reads (we take into account that a pair of $V$ lines crosses at higher-order van Hove X point, so that  to avoid doubling the  contribution of X point vicinity we divide this by  factor of 2)
\begin{equation}
    \Delta\rho_V(\epsilon, \tau = 0) = 3\int_{\mathcal{O}(V)} \frac{d^3\mathbf{k}}{V_{\rm BZ}}\delta(\epsilon - t_{\rm fcc}\left(\mathbf{k};\tau = 0\right)),
\end{equation}
 where the Brillouin zone volume for fcc lattice reads $V_{\rm BZ} = 4\pi^3$. $q_x = (1-\cos k_z)^{-1/2}q_\perp\cos\varphi$, $q_y = (1+\cos k_z)^{-1/2}q_\perp\sin\varphi$ and $\mathcal{O}(V)$ is determined by cutoff conditions. We obtain 
 \begin{multline}\label{eq:delta_rho.tau=0}
    \Delta\rho_{V}(\epsilon, \tau = 0)  
    \\= \frac3{V_{\rm BZ}}\int\limits_{-\pi/2}^{\pi/2} dk_z \int dq_xdq_y
    \delta\left(\epsilon - E_V(k_z,q_x,q_y)\right)  
    \\
    = \frac3{V_{\rm BZ}}\int\limits_{-\pi/2}^{\pi/2} \frac{dk_z}{|\sin k_z|} \int\limits_0^{\bar{q}_\perp\left(k_z\right)} dq_\perp q_\perp
    \int\limits_0^{2\pi} d\varphi\delta\left(\epsilon - 4 +  2q^2_\perp \right.  \\
   \left. - \frac{q^4_\perp}{12}\left(\frac{\cos^4\varphi}{\sin^2\frac{k_z}2}  + \frac{\sin^4\varphi}{\cos^2\frac{k_z}2} + \frac{3\sin^22\varphi}{\sin^2k_z}\right)\right),  
\end{multline}
%??higher order vHs??
where $k_z$-dependent cutoff parameter is introduced which is determined from fourth-order terms in the spectrum expansion. % (\ref{eq:t_fcc_tau=0_expansion}).
Main contribution to the integral originates from small $k_z$, so that $\sin^4\varphi$ term is irrelevant. Maximal value of the fourth-order term is ${q^4_\perp}/(3k_z^2)$. From the condition $2q_\perp^2 = {q^4_\perp}/(3k_z^2)$ we find an estimation of the scale where fourth-order terms become important, $\bar{q}_\perp\left(k_z\right) = \alpha_Vk_z$, $\alpha_V \sim \sqrt{6}$.  
%Neglecting the fourth-order terms yields
 %\begin{equation}
 %   \Delta\rho_{V}(\epsilon, \tau = 1)  
 %   = \frac{3\cdot4\pi}{V_{\rm BZ}}\int\limits_0^{\pi/2} \frac{dk_z}{\sin k_z} \int\limits_0^{\bar{q}_\perp\left(k_z\right)} dq_\perp q_\perp
 %   \delta\left(\epsilon + 4 -  2q^2_\perp \right)
%\end{equation}
%Or
%\begin{equation}
%    \Delta\rho_{\rm fcc}(\epsilon, \tau = 1)  
%    = \frac{3\pi}{V_{\rm BZ}}\theta\left(\epsilon + 4 \right)\int\limits_0^{\pi/2} \frac{dk_z}{\sin k_z}
%    \theta\left(2\bar{q}^2_\perp\left(k_z\right) - \epsilon - 4 \right),
%\end{equation}
%or
%\begin{equation}
%    \Delta\rho_{\rm fcc}(\epsilon, \tau = 1)  
%    = -\frac{3\pi}{V_{\rm BZ}}\theta\left(\epsilon + 4 \right)\ln\tan\sqrt{\frac{\epsilon+4}{8\alpha^2_V}},
%\end{equation}
%so 
We get finally the asymptotics valid at small $4 - \epsilon$:
\begin{equation}
    \Delta\rho_{V}(\epsilon, \tau = 0)  = \frac3{8\pi^2}\theta\left(4 - \epsilon \right)\ln\frac{8\alpha^2_V}{4 - \epsilon}.
\end{equation}

The line $\varLambda$, joining van Hove points $\Gamma$ and L, becomes van Hove line at $\tau = 1$.  We consider this case   by expanding spectrum in the vicinity of van Hove line $\varLambda$: $k_x = k_y = k_z = k_\varLambda$, $0<k_\lambda<\pi/2$. %It is convenient to choose (local) orthonormal basis $\mathbf{e}_1(+1/\sqrt{2},0,-1/\sqrt{2})$, $\mathbf{e}_2(-1/\sqrt{6},+2/\sqrt{6},-1/\sqrt{6})$, $\mathbf{e}_2(+1/\sqrt{3},+1/\sqrt{3},+1/\sqrt{3})$. 
We parametrize the line $\Lambda$ as $\mathbf{k}_\varLambda(q_1, q_2, k_\varLambda) = q_\perp\cos\varphi\mathbf{e}_1 + q_\perp\sin\varphi\mathbf{e}_2 + \sqrt{3}k_\varLambda\mathbf{e}_3$, where the orthonormal basis $\mathbf{e}_1(+1/\sqrt{2},0,-1/\sqrt{2})$, $\mathbf{e}_2(-1/\sqrt{6},+2/\sqrt{6},-1/\sqrt{6})$, $\mathbf{e}_2(+1/\sqrt{3},+1/\sqrt{3},+1/\sqrt{3})$ was used. We have up to  higher order terms
%\begin{multline}
%    t_{\rm fcc}\left(\mathbf{k}_\varLambda(q_1, q_2, k_\varLambda); \tau = 1\right) = 6 - 6\sin^2 k_\varLambda(q_1^2 + q_2^2) \\
%    + \sqrt{6}\sin k_\varLambda\cos k_\varLambda (3q_1^2 - q_2^2)q_2 + \frac{3 - 5\cos 2k_\varLambda}8(q_1^2 + q_2^2)^2.
%\end{multline}
%We use representation $q_1 = q_\perp\cos\varphi$, $q_2 = q_\perp\sin\varphi$. 
%\begin{multline}\label{eq:tau=1.expansion}
$E_\varLambda(k_\varLambda,q_\perp, \varphi)=t_{\rm fcc}\left(\mathbf{k}_\varLambda(q_\perp, \varphi, k_\varLambda); \tau = 1\right) = -6 + 6\sin^2 k_\varLambda q_\perp^2 
    - \sqrt{6}\sin k_\varLambda\cos k_\varLambda \sin\varphi(3\cos^2\varphi - \sin^2\varphi) q^3_\perp - (3 - 5\cos 2k_\varLambda)q^4_\perp/8$.
%\end{multline}
Since in the vicinity of $k_\varLambda = 0$ the leading non-constant contribution starts from fourth-order terms in $q_\perp$ and $k_z$, we see that $\Gamma$ is \textit{a van Hove higher-order point} with all the inverse masses being zero.  
%Maximal absolute value of the factor $\sin\varphi(3\cos^2\varphi - \sin^2\varphi)$ is 1. So we obtain upper bound for the third and fourth order terms: $M_3(k_\varLambda) = \sqrt{6}\sin k_\varLambda\cos k_\varLambda q^3_\perp$, $M_4(k_\varLambda) = \frac{3 - 5\cos 2k_\varLambda}8q^4_\perp$. We get upper cutoff $6\sin^2 k_\varLambda q_{\perp,3}^2 \sim M_3(k_\varLambda)$ therefore $q_{\perp,3} = \sqrt{6}\tan k_\varLambda \sim \sqrt{6}k_\varLambda$. Analogously $6\sin^2 k_\varLambda q_{\perp,4}^2\sim\frac{5\cos 2k_\varLambda - 3}8q^4_{\perp,4}$ therefore $q_{\perp,4} = \frac{4\sqrt{3}\sin k_\varLambda}{\sqrt{5\cos 2k_\varLambda - 3}}\sim 2\sqrt{6}k_\varLambda$. 
Analogously to the case of $V$ line we introduce the cutoff for $\bar{q}_{\perp} = \alpha_\perp k_\varLambda$, where numerical value $\alpha_\perp \sim \sqrt6$ and should be determined more precisely later. 

\begin{figure}[t!]
\noindent\includegraphics[angle=-90,width=0.45\textwidth]{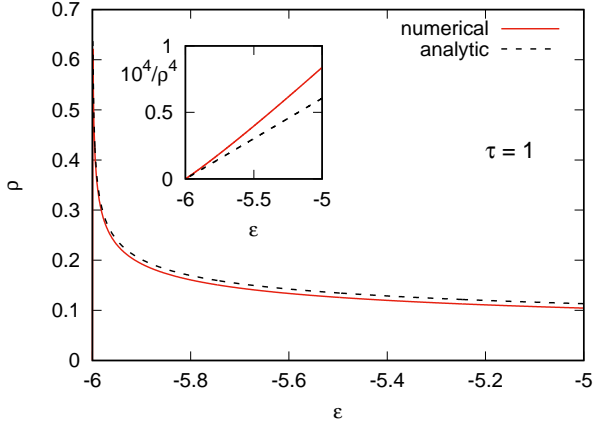}
\caption{$\tau = 1$. Density of states $\rho_{\rm fcc}(\epsilon, \tau = 1)$ (see below Eq.~(\ref{eq:rho_zeta_final_mod})) and its analytic ansatz  $\Delta\rho_{\varLambda}(\epsilon, \tau = 1)$, see Eq.~(\ref{eq:delta_rho.tau=1_asymp}), in the vicinity of band bottom. In the inset the plots of $1/\rho^4_{\rm fcc}(\epsilon, \tau = 1)$ and  $1/\Delta\rho^4_{\varLambda}(\epsilon, \tau = 1)$ are shown. }
\label{fig:dos_tau=1_asymp}
\end{figure}

We extract from the definition (\ref{eq:DOS_fcc_def}) the contribution from the vicinity eight $\Lambda$ lines 
%\begin{equation}\label{eq:delta_rho.tau=1.start}
%    $\Delta\rho_{\varLambda}(\epsilon, \tau = 1) = 8\int_{\mathcal{O}(\varLambda)} \frac{d^3\mathbf{k}}{V_{\rm BZ}}\delta(\epsilon - t_{\rm fcc}\left(\mathbf{k};\tau = 1\right))$. 
%\end{equation}
%where the spectrum is determined by Eq.~(\ref{eq:spectrum}) and the presence of equivalent by symmetry 8 van Hove line is taken into account. 
\begin{multline}\label{eq:delta_rho.tau=1}
    \Delta\rho_{\varLambda}(\epsilon, \tau = 1) = 8\int_{\mathcal{O}(\varLambda)} \frac{d^3\mathbf{k}}{V_{\rm BZ}}\delta(\epsilon - t_{\rm fcc}\left(\mathbf{k};\tau = 1\right))\\ 
    = \frac8{4\pi^3}\int\limits_0^{\pi/2} dk_\varLambda \int\limits_0^{\bar{q}\left(k_\varLambda\right)} dq_\perp q_\perp
    \int\limits_0^{2\pi} d\varphi\delta\left(\epsilon - E_\varLambda(k_\varLambda,q_\perp, \varphi)\right).  
\end{multline}
Due to the use of the cutoff, we can neglect subleading terms in formula (\ref{eq:delta_rho.tau=1}) to obtain:
%\begin{equation}\label{eq:delta_rho.tau=1.2}
%    \Delta\rho_{\varLambda}(\epsilon, \tau = 1) = \frac4{\pi^2}\int\limits_0^{\pi/2} dk_\varLambda \int\limits_0^{\bar{q}\left(k_\varLambda\right)} dq_\perp q_\perp
%   \delta\left(\epsilon - 6 + 6\sin^2 k_\varLambda q_\perp^2\right).  
%\end{equation}
%\begin{equation}\label{eq:delta_rho.tau=1_asymp}
%    \Delta\rho_{\varLambda}(\epsilon, \tau = 1) = \frac1{3\pi^2}\int\limits_0^{\pi/2} \frac{dk_\varLambda}{\sin^2k_\varLambda}\theta\left(\epsilon - 6 + 6\sin^2 k_\varLambda \bar{q}_\perp^2(k_\varLambda)\right).  
%\end{equation}
%If we assue that $\epsilon$ is slightly deviates from $\epsilon = 6$ we replace $\sin k_\varLambda$ by $k_\varLambda$ and get
\begin{equation}\label{eq:delta_rho.tau=1_asymp}
    \Delta\rho_{\varLambda}(\epsilon, \tau = 1) = %\frac1{3\pi^2} \cot {\sqrt[4]{\frac{6 - \epsilon}{6\alpha^2_\perp}}} \approx 
    \frac1{3\pi^2}\left(\frac{6\alpha^2_\perp}{6 + \epsilon}\right)^{1/4}.  
\end{equation}
To get a close agreement for Eq.~(\ref{eq:delta_rho.tau=1_asymp}) and numerical calculation (see below) in the vicinity $\epsilon = -6$ point we choose $\alpha_\perp = 1.88\sqrt6 \approx 4.6$. 
We compare $\rho_{\rm fcc}(\epsilon, \tau = 1)$ calculated by direct using Eq.~(\ref{eq:rho_zeta_final_mod}), see below, and asymptotics~(\ref{eq:delta_rho.tau=1_asymp}) with a chosen $\alpha_\perp = 4.6$ in Fig.~\ref{fig:dos_tau=1_asymp}. 
One case see a good agreement in the vicinity of the vHS point at $\epsilon = -6$.

\section{Fully analytical treatment of the density of states}

To derive an analytic expression for the DOS we use  the effective  one-dimensional presentation of the spectrum (\ref{eq:spectrum}):
\begin{multline}
t_{\rm fcc}\left(\mathbf{k};\tau\right) = t_0(k_x, k_y) 
- 2p(k_x,k_y)\cos k_z - 2p'(1+\cos 2k_z),
\end{multline}
where 
\begin{eqnarray}
\label{eq:t0_def}
	\nonumber
	t_0(k_x,k_y) &=& -4t\cos k_x\cos k_y 
	 - 2\tau t(1- \cos 2k_x - \cos 2k_y) ,
\label{eq:t_def}\\
	p(k_x,k_y) &=& 2t(\cos k_x + \cos k_y),
\end{eqnarray}
and $p' = -\tau t$. Below we set for brevity $t = 1$. 
Investigation of the spectrum (\ref{eq:spectrum})~\cite{2019:Igoshev-JETP} yields the following boundaries for it
\begin{eqnarray}
\nonumber
    E_{\rm min}(\tau) &=& {\rm min}[-12 + 6\tau, -6\tau],\\
    E_{\rm max}(\tau) &=& {\rm max}[4 + 6\tau, 4 + 2\tau, -6\tau].    
\end{eqnarray}

We directly use the definition of the density of states
\begin{multline}\label{eq:DOS_fcc_treatment_start}
	\rho_{\rm fcc}(\epsilon;\tau) = \iiint\limits_{0<k_i<\pi}\frac{dk_x}{\pi}\frac{dk_y}{\pi}\frac{dk_z}{\pi}\delta(\epsilon - t_{\rm fcc}\left(\mathbf{k};\tau\right)).  
	%\\
	%= \iiint\limits_{0<k_i<\pi}\frac{dk_x}{\pi}\frac{dk_y}{\pi}\frac{dk_z}{\pi}\delta( t_0(k_x,k_y) - 2t_{1D}'(k_x,k_y) - 2t_{1D}(k_x,k_y)\cos k_z - 2t_{1D}'(k_x,k_y)\cos 2k_z - \epsilon )	
\end{multline}
to obtain
\begin{equation}\label{eq:rho_fcc_through_rho_1d}
\rho_{\rm fcc}(\epsilon) = \iint\limits_{0<k_i<\pi}\frac{dk_x}{\pi}\frac{dk_y}{\pi}\rho_{1D}(\epsilon - t_0(k_x,k_y), t_{1D}(k_x,k_y), t'_{1D}),
\end{equation}
where we have  introduced the known DOS 
\begin{equation}\label{eq:DOS_1D_formal_def}
\rho_{1D}(\epsilon; p,p') = \int\limits_0^\pi\frac{dk}{\pi}\delta(\epsilon - t_{1D}(p,p'))
\end{equation}
for the 1D spectrum within nearest and next-nearest neighbor hopping approximation (with effective integrals $p$ and $p'$)
\begin{equation}
t_{1D}(k;p,p') =  - 2p\cos k -  2p'(1+\cos 2k).
\end{equation}
An explicit expression for (\ref{eq:DOS_1D_formal_def}) is
\begin{equation}\label{eq:DOS_1D}
\rho_{1D}(\epsilon; p,p') = \sum_{\sigma=\pm1} \bar{\rho}_{1D}\left(\epsilon; \sigma |p|,p'\right), 
\end{equation}
where
\begin{multline}\label{eq:DOS_1D_bar}
\bar{\rho}_{1D}\left(\epsilon; p,p'\right) = \frac1{2\pi} \sqrt{\frac{\theta(p^2 - 4p'\epsilon)}{p^2 - 4p'\epsilon}}
\varphi\left(\frac{\sqrt{p^2 - 4p'\epsilon} + p}{4p'}\right)
\end{multline}
and
\begin{equation}\label{eq:phi_def}
\varphi(x) = g(1-x^2)
\end{equation}
with $g(x) = \sqrt{\theta(x)/x}$ and $\theta(x)$ being the Heaviside step function.

%Substituting the expressions (\ref{eq:t0_def}), (\ref{eq:t_def}) and (\ref{eq:t'_def}) into the Eq.~(\ref{eq:DOS_1D})
%\begin{equation}\label{eq:DOS_fcc1}
%\rho_{\rm fcc}(\epsilon, \tau) = \sum_{\sigma=\pm1}\int\limits_{0}^\pi\int\limits_{0}^\pi\frac{dk_x}{\pi}\frac{dk_y}{\pi}\frac1{2\pi} \sqrt{\frac{\theta(\mathcal{D}(k_x, k_y))}{\mathcal{D}(k_x, k_y)}}\varphi\left(\frac{\sqrt{\mathcal{D}(k_x, k_y)} + \sigma t_{1D}(k_x,k_y)}{4t'_{1D}}\right),
%\end{equation}
%where the ``Discriminant''
%\begin{equation}
%\mathcal{D}(k_x, k_y) = t^2_{1D}(k_x,k_y) - 4t'_{1D} (\epsilon - t_0(k_x,k_y))
%\end{equation}
%is introduced for brevity. 
Further simplification can be achieved by  introducing symmetrized variables: $s(k_x, k_y) = (1/2)(\cos k_x + \cos k_y)$, $\xi(k_x, k_y) = (1/2)|\cos k_x - \cos k_y|$.
We thereby obtain
\begin{multline}\label{fun:rho} % with Jelitto definition of energy e!!!
\rho_{\rm fcc}(\epsilon, \tau) = \frac1{\pi^{3}}\int\limits_{-1}^{+1}ds\int\limits_{0}^{1 - |s|} \frac{d\xi}{\sqrt{((1 - |s|)^2 - \xi^2)((1 + |s|)^2 - \xi^2)}}\\
\times\varphi\left(\left(1/\mathcal{K}(\epsilon, \tau, s, \xi^2) + s\right)/\tau\right)\mathcal{K}(\epsilon, \tau, s, \xi^2),
\end{multline}
where
\begin{equation}
\label{eq:R_def}
	\mathcal{K}(\epsilon, \tau, s, a) = g\left(1 + \tau - 2\tau^2)s^2 + \tau ((\epsilon -  6\tau)/4 - (1 + 2\tau)a)\right)
\end{equation}
is introduced for brevity. 
We rescale $\xi^2 = (1 - |s|)^2\zeta$ to make inner integration limits constant:
%\begin{equation}\label{eq:rho_zeta_final}
%\rho_{\rm fcc}(\epsilon, \tau) = \frac1{\pi^{3}}\int\limits_{-1}^{+1}ds\int\limits_{0}^{1}\frac{d\zeta}{\sqrt{1-\zeta^2}} \frac{\varphi\left(\frac{\sqrt{\mathcal{R}(\epsilon_1, \tau, s, \zeta)} + s}{\tau}\right)}{\sqrt{(1 + |s|)^2 - (1-|s|)^2\zeta^2}\sqrt{\mathcal{R}(\epsilon_1, \tau, s, \zeta)}}
%\end{equation}
%let's introduce one more variable change for convenience
\begin{equation}\label{eq:rho_zeta_final_mod}
\rho_{\rm fcc}(\epsilon, \tau) = \frac1{2\pi^{3}}\int\limits_{-1}^{+1}ds\int\limits_{0}^{1} \frac{d\zeta  Q( \epsilon,\tau , s , \zeta)}{\sqrt{(1-\zeta)\zeta}},
\end{equation}
\begin{equation}\label{eq:Q_def}
    Q( \epsilon,\tau , s , \zeta) =  \frac{\varphi\left(\left(1/\mathcal{K}(\epsilon, \tau, s, (1 - |s|)^2\zeta) + s\right)/\tau\right)\mathcal{K}(\epsilon, \tau, s, (1 - |s|)^2\zeta))}{\sqrt{(1 + |s|)^2 - (1-|s|)^2\zeta}}.
\end{equation}
%and ``$R$-discriminant'' 
%\label{eq:R_def}
%	\mathcal{R}(\epsilon, \tau, s, \zeta) = (1 + \tau - 2\tau^2)s^2 - \tau ((\epsilon -  6\tau)/4 + (1 + 2\tau)(1-|s|)^2\zeta)
%\end{equation}
%or
%\begin{equation}
%	\mathcal{R}(\epsilon_1, \tau, s, \zeta) = (1 - \tau + 2\tau^2)s^2+ \tau(1 + 2\tau)(1-|s|)^2 \left(\frac{\epsilon_1}{4(1 + 2\tau)(1-|s|)^2} + \zeta^2\right)
%\end{equation}
The formulas (\ref{eq:phi_def}), (\ref{eq:R_def}--\ref{eq:Q_def}) provide analytical  results which are convenient for numerical calculations.
%Actually $\mathcal{D} = 16\mathcal{R}$ in terms of other variables. 

\section{Comparison with tetrahedron method}
\begin{figure}[t!]
\noindent
\includegraphics[angle = -90, width=0.45\textwidth]{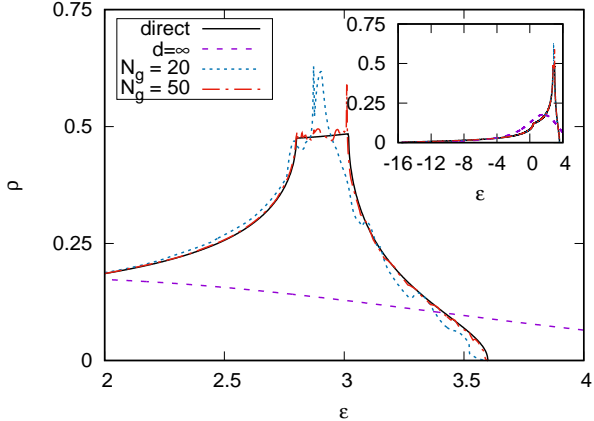}
\caption{Comparison of DOS for fcc lattice with $\tau = -0.6$ in the vicinity of plateau calculated directly using Eq.~(\ref{eq:rho_zeta_final_mod}), tetrahedron method with tetrahedron number $6\times N_{\rm mesh}\times N_{\rm mesh}\times N_{\rm mesh}$, for the grid resolution $N_{\rm g} = 20, 50$ and in the limit $d = \infty$, see derivation in the Appendix. The inset shows the whole bandwidth energy range. 
\label{fig:tetra_comparison}
}
\end{figure}

A commonly used tetrahedron method~\cite{1994:Andersen} is a popular tool of density-of-states calculation within density functional theory~\cite{1989:Jones}. A fundamental block of this approach is linear approximation of the spectrum within an elementary tetrahedron. However, when one of tetrahedron vertices is a van Hove point, its contribution, being actually dominating, is crudely averaged with the contributions of another vertices. This implies very bad accuracy of the method when the energy is in the vicinity of van Hove points. 
In~Fig.~\ref{fig:tetra_comparison}, a comparison of the results of  direct calculation using Eq.~(\ref{eq:rho_zeta_final_mod}) and tetrahedron method is shown. For the latter method, we find strong  artificial oscillations in the vicinity of van Hove energies. The convergence of the results in the limit $N_{\rm mesh}\rightarrow\infty$ is thereby substantially poor. The origin of this discrepancy is an averaging of vertices of particular tetrahedron with zero velocity and non-zero on equal foot, which underestimates the contribution of van Hove singularity point vicinity in reciprocal space, see also~Ref.~\cite{2017:Stepanenko}. 

\section{Conclusions}
To conclude,  we have analyzed in detail the van Hove singularities in the excitation spectrum of fcc lattice. These  singularities  can play an important role in the temperature dependencies of physical properties. e.g.,  electronic heat capacity, magnetic susceptibility, thermoelectric power (for bipartite lattices, these effects were treated in Refs.\cite{2019:Igoshev-PMM,2019:Igoshev-JETP}).
In this connection, examples of fcc Ni and ZrZn$_2$ (the C15 Laves phase crystal structure which is a superposition of  fcc sublattices) can be once more mentioned. The vHS can be also important for some lattice and magnetic  characteristics owing to anomalies of phonon and magnon spectra. 
Especial role belongs to van Hove lines 
intersection typically forming a higher-order van Hove point, which leads to more strong effects~\cite{2021:Hung}. %Formally, we have here a situation of higher-order van Hove points which typically provided as van Hove lines intersections. 

We also state that DOS
%in energy regions 
in the vicinity of vHS energies can be poorly captured by standard approaches 
%like conventional in 
within dynamical mean-field theory ($d=\infty$ limit) and tetrahedron method, which are  popular in~density-functional calculations. 
Our analysis can be useful to supplement computational approaches which deal with  the electron spectrum and properties of actual materials.

The research was carried out within the state assignment of the Ministry of Science and Higher Education of the Russian Federation (theme  ``Quantum''  No. 122021000038-7) and supported in part by
Russian Foundation for Basic Research (project No. 20-02-00252). 
The calculations were performed on the Uran supercomputer at the IMM UB RAS. 

\section*{Appendix. $d=\infty$ case}
In~Ref.~\cite{Ulmke}, the  fcc-lattice DOS in the limit of infinite space dimension $d$ was considered in the special case $t ' = t/2$ (where giant quasi-one-dimension van Hove singularity  is formed owing to van Hove surface). Passing to such a limit may be useful in the context of dynamical mean-field approach implementation. Therefore, it is instructive to compare the DOS form in the limit of infinite dimension with the case $d = 3$. To derive an explicit expression for DOS in this limit we have to set the scaling of hopping integrals with $d$: $t = t_1/\sqrt{d (d - 1)}$, $t' = t_2/\sqrt{2d}$ (this is based on the fact that a site in fcc lattice has $2d(d-1)$ nearest and $2d$ next-nearest neighbours). The derivation uses ``statistical'' independence of the quantities $\xi = \sqrt{2/d}\sum_i \cos k_i$ and $\eta = \sqrt{2/d}\sum_i \cos 2k_i$ in the limit $d\rightarrow\infty$. Since $\varepsilon_{\bf k} = -t_1(\xi^2 - 1) + t_2 \eta$, we obtain directly
\begin{equation}\label{eq:DOS_at_d=infty}
	\rho_{d=\infty}(\varepsilon; t_1, t_2) = \frac {\exp\left[-(\varepsilon-t_1)^2/(2t^2_2)\right]}{2\pi\sqrt{|t_1t_2|}}f\left(\frac{t^2_2-2t_1(\varepsilon - t_1)}{|t_1t_2|}\right),
\end{equation}
where
\begin{equation}
f(x) = \frac{e^{x^2/16}}2\sqrt{|x|}
\begin{cases}
K_{1/4}(x^2/16),& x > 0,\\
\frac{\pi}{\sqrt2}\left(I_{1/4}(x^2/16)+I_{-1/4}(x^2/16)\right),& x < 0,
\end{cases}
\end{equation}
where $I_\nu$ and $K_\nu$ are Infeld and McDonald functions, 
%with $\nu = 1/4$ 
respectively. 

The comparison of the results for the cases $d = 3$ and $d \rightarrow \infty$ at $\tau = -0.6$ is shown in Fig.~\ref{fig:tetra_comparison}. We can see that the $d = 3$ DOS possesses pronounced van Hove plateau, which is destroyed in the limit $d \rightarrow \infty$. Note also that the case of giant van Hove singularity $t' = t/2$ is not captured by Eq.~(\ref{eq:DOS_at_d=infty}). 

%\bibliography{mybibfile}

\begin{thebibliography}{99}

%Классическая работа ван Хова, теорема ван Хова!!!
% The Occurrence of Singularities in the Elastic Frequency Distribution of a Crystal
\bibitem{1953:vanHove} L. Van Hove, Phys. Rev.~\textbf{89}, 1189 (1953).

% Localized and itinerant behavior of electrons in metals
\bibitem{1993:Vonsovskii} S. V. Vonsovskii, M. I. Katsnelson, and A. V. Trefilov, Phys. Met. Metallogr. \textbf{76}, 247 (1993).

% Collective electron ferromagnetism
\bibitem{1938:Stoner} E.~C.~Stoner, Proc.~Roy.~Soc.~A~\textbf{165}, 372~(1938).

% Magnetic Isotherms in the Band Model of Ferromagnetism
\bibitem{1968:Edwards} D. M. Edwards and E. P. Wohlfarth, Proc. R. Soc. Lond. A  \textbf{303}, 127~(1968).

%Importance of the van Hove singularity in superconducting PdTe2
\bibitem{Park}
K. Kim, S. Kim, J. S. Kim, H. Kim, J.-H. Park, and B. I. Min, 
Phys. Rev. B \textbf{97}, 165102 (2018).
%The singularities in the density of electron states and their effect on the elasticity moduli in alkaline earth metals
\bibitem{1990:Peschanskikh}  M. I. Katsnel’son, G. V. Peschanskikh, and A. V. Trefilov, Sov. Phys. Solid State \textbf{32}, 272 (1990).

% Magic of high-order van Hove singularity
\bibitem{2019:Yuan} N.~F.~Q. Yuan, H. Isobe, L. Fu, Nat. Comm.~\textbf{10}, 5769~(2019).

% Unconventional superconductivity and density waves in twisted bilayer graphene
\bibitem{2018:Isobe} H.~Isobe, N.~F.~Q.~Yuan, L.~Fu,  Phys.~Rev.~X~\textbf{8}, 041041 (2018).

%Nematic superconductivity stabilized by density wave fluctuations: application to twisted bilayer graphene
\bibitem{2019:Kozii} V.~Kozii, H.~Isobe, J.~W.~F.~Venderbos, L.~Fu, Phys. Rev. B \textbf{99}, 144507 (2019).


% Enhanced Thermoelectric Performance of Type-II Nodal-Line Semimetals by van Hove Singularities in Density of States 
% About Mg3Bi2
\bibitem{2021:Hung} N.~T.~Hung, J.~M. Adhidewata, A.~R.~T. Nugraha, R.~Saito, Phys. Rev.~B~\textbf{105}, 115142 (2022). 
% Electronic correlations in nodal-line semimetals
% About ZrSiSe
\bibitem{2020:Shao}  Y. Shao, A. N. Rudenko, J. Hu, Z. Sun, Y. Zhu, S. Moon, A. Millis, S. Y., A. I. Lichtenstein, D. Smirnov, Z. Q. Mao, M. I. Katsnelson, and D. N. Basov, Nat. Phys. \textbf{16}, 636 (2020).

% Excitonic Instability and Pseudogap Formation in Nodal Line Semimetal ZrSiS
% About ZrSiS
\bibitem{2018:Rudenko} A. N. Rudenko, E. A. Stepanov, A. I. Lichtenstein, and M. I. Katsnelson, Phys. Rev. Lett. \textbf{120}, 216401 (2018).

%Effect of density of states peculiarities on Hund’s metal behavior
%Destription: About density of states peculiarities on QP damping in Hund's metals
\bibitem{2018:Belozerov} A.~S.~Belozerov, A.~A.~Katanin and V.~I.~Anisimov, Phys.~Rev.~B~\textbf{97}, 115141 (2018).

% Orbital-selective Mott-insulator transition in Ca2−x SrxRuO4
\bibitem{2002:Anisimov} V. I. Anisimov, I. A. Nekrasov, D. E. Kondakov, T. M. Rice, and M. Sigrist, Eur. Phys.~J.~B~\textbf{25}, 191 (2002).

%Linear-Temperature Dependence of Static Magnetic Susceptibility in LaFeAsO from Dynamical Mean-Field Theory
\bibitem{2011:Skornyakov}  S.~L.~Skornyakov, A.~A.~Katanin, and V.~I.~Anisimov, Phys. Rev. Lett.~\textbf{106}, 047007~(2011). 

%The density of states of some simple excitations in solids
\bibitem{1969:Jelitto}  R. J. Jelitto,  J. Phys. Chem. Solids 30, 609 (1969).


\bibitem{2010:Igoshev} P.A.~Igoshev, M.A.~Timirgazin, A.A.~Katanin, A.K.~Arzhnikov and V.Yu.~Irkhin, Phys. Rev. B \textbf{81}, 094407 (2010).

\bibitem{2015:Igoshev} P.A.~Igoshev, M.A.~Timirgazin, V.F.~Gilmutdinov, A.K.~Arzhnikov and V. Yu.~Irkhin, J. Phys.: Cond. Matt. \textbf{27}, 446002 (2015).

\bibitem{2003:Kampf} A. A. Katanin and A. P. Kampf, Phys. Rev. B \textbf{68}, 195101 (2003).

\bibitem{Hlubina} R. Hlubina, S. Sorella, and F. Guinea, Phys. Rev. Lett. \textbf{78}, 1343 (1997); R. Hlubina, Phys. Rev. B \textbf{59}, 13960 (1999).

%%%% ZrZn2 Refs. %%%%
%Heavy Quasiparticles in the Ferromagnetic Superconductor ZrZn2
\bibitem{2003:ZrZn2:Yates} S. J. C. Yates, G. Santi, S. M. Hayden, P. J. Meeson, and S. B. Dugdale, Phys.~Rev.~Lett.~\textbf{90}, 057003~(2003).

% Direct Observation of the Multisheet Fermi Surface in the Strongly Correlated Transition Metal Compound ZrZn2
\bibitem{2004:ZrZn2:Major} Zs. Major, S. B. Dugdale, R. J. Watts, G. Santi, M. A. Alam, S. M. Hayden, J. A. Duffy, J.W. Taylor, T. Jarlborg, E. Bruno, D. Benea, and H. Ebert, Phys. Rev. Lett. \textbf{92}, 107003 (2004).

% Quantum Metamagnetic Transitions Induced by Changes in Fermi-Surface Topology: Applications to a Weak Itinerant-Electron Ferromagnet ZrZn2
\bibitem{2007:ZrZn2:Amaji} Y. Amaji , T. Misawa, M. Imada, J. Phys. Soc. Jpn \textbf{76}, 063702 (2007).

%Non-Fermi Liquid State Bounded by a Possible Electronic Topological Transition in ZrZn2
\bibitem{2012:ZrZn2:Kabeya} N. Kabeya, H. Markawa, K. Deguchi, N. Kimura, H. Aoki, N.~K.~Sato, J. Phys. Soc. Jpn  \textbf{81}, 073706~(2012).
 
%%%%%%%%%%%%%%%%%%%%%%%%%
% Local magnetic moments in iron and nickel at ambient and Earth’s core conditions
% https://doi.org/10.1038/ncomms16062
\bibitem{2017:Ni:vanHoveMagnet} A.~Hausoel, M.~Karolak, E.~Sasioglu, et al. Nat. Commun.~\textbf{8}, 16062 (2017). 

%Giant Van Hove Density of States Singularities and Anomalies of Electron and Magnetic Properties in Cubic Lattices
\bibitem{2019:Igoshev-PMM} P.~A.~Igoshev and V.~Yu.~Irkhin, Phys.  Met. Metallogr.~\textbf{120}, 1282~(2019).

%Electron Spectrum Topology and Giant Singularities of the Electron Density of States in Cubic Lattices
\bibitem{2019:Igoshev-JETP} P.~A.~Igoshev and V.~Yu.~Irkhin, JETP Lett.~\textbf{110}, 727~(2019).

%Green’s functions of the face-centered-cubic Heisenberg ferromagnet with second-neighbor interactions
\bibitem {1972:Swendsen} R. H. Swendsen and H. Callen,  Phys. Rev. B \textbf{6}, 2860 (1972).

% Ferromagnetism in the Hubbard model on fcc-type lattices

\bibitem {Ulmke}  M. Ulmke,  Eur. Phys. J. B \textbf{1}, 301 (1998).

\bibitem{2017:Igoshev} P.A.~Igoshev, E.E.~Kokorina, I.A.~Nekrasov, Physics of Metals and Metallography~\textbf{118}, 207~(2017).

% Tetrahedron method
\bibitem{1994:Andersen}
P.~E.~Bl\"ochl, O.~Jepsen, O.~K.~Andersen, Phys.~Rev.~B~\textbf{49}, 16223 (1994).

\bibitem{1989:Jones}
R.O.~Jones, O.~Gunnarsson, Rev. Mod. Phys.~\textbf{61}, 689 (1989).

% Kohn anomalies in momentum dependence of magnetic  susceptibility of some three-dimensional systems
\bibitem{2017:Stepanenko} A.A. Stepanenko, D.O. Volkova, P.A. Igoshev, A.A. Katanin, J. Exp. Theor. Phys.
\textbf{125}, 879~(2017).


\end{thebibliography}

\end{document}